\documentclass[acmsmall,nonacm]{acmart}
\usepackage{listings}

\lstdefinelanguage{Rust}{
  morekeywords={
    as,break,const,continue,crate,else,enum,extern,false,fn,for,
    if,impl,in,let,loop,match,mod,move,mut,pub,ref,return,self,
    Self,static,struct,super,trait,true,type,unsafe,use,where,while,
    async,await,dyn
  },
  sensitive=true,
  morecomment=[l]{//},
  morecomment=[s]{/*}{*/},
  morestring=[b]",
  morestring=[b]'
}

\usepackage{booktabs}
\usepackage{tabularx}
\usepackage{array}
\usepackage{multirow}
\usepackage{makecell}

\AtBeginDocument{%
  }

\title{XRust: Detecting Logical Races in Rust Linux Drivers}
\author{Qingxiao Xu}
\affiliation{\institution{Texas A\&M University}\country{USA}}
\author{Sheng Ze}
\affiliation{\institution{Texas A\&M University}\country{USA}}
\author{Zhicheng Chen}
\affiliation{\institution{Texas A\&M University}\country{USA}}
\author{Arjun Manikandan}
\affiliation{\institution{Texas A\&M University}\country{USA}}
\author{Jeff Huang}
\affiliation{\institution{Texas A\&M University}\country{USA}}

\begin{document}

\renewcommand{\shortauthors}{Xu et al.}

\begin{abstract}
Rust is increasingly used in Linux to improve memory safety, but its guarantees do not prevent logical races: individually synchronized operations can still interleave in ways that violate higher-level invariants. Detecting such bugs typically requires manually written specifications and carefully controlled execution environments.

We present XRust, a technique for detecting logical races in Linux Rust drivers without manually specified invariants. Using Logical Happens-Before (LHB) analysis, XRust identifies shared-state accesses relevant to logical ordering, automatically constructs operation-pair harnesses, and uses serial differential oracles to detect concurrent executions inconsistent with either serial order. XRust also uses lightweight userspace abstractions to execute driver code under reproducible schedules. Applied to Linux 7.3-rc2, XRust identifies 17 shared objects across 102 Rust files from ten drivers, constructs 1,654 executable harnesses, and discovers two previously unknown logical races. The rnull block-driver race has been confirmed by Linux maintainers, with a fix accepted into the Linux block subsystem tree.

\end{abstract}



\maketitle

\section{Introduction}


Rust offers a path toward eliminating memory-safety vulnerabilities and data races through its ownership and type systems~\cite{rust_book}. These guarantees are particularly compelling for operating-system kernels, where a single defect can have system-wide consequences~\cite{rfl_empirical_2024}. 
Linux, historically implemented primarily in C, has long faced memory-safety and concurrency bugs due to manual memory management, unrestricted pointer manipulation, and complex synchronization protocols~\cite{isrg_memory_safety_2025}. 

Rust-for-Linux (RFL), launched in 2020, is an effort to bring Rust's memory- and thread-safety guarantees into the Linux kernel and enable kernel components, particularly device drivers, to be written in Rust~\cite{corbet_rust_linux_2022, linux_rust_docs}. The first functional Rust driver was merged into Linux 6.8 in early 2024. Since then, Rust adoption in Linux has accelerated, growing from 95 SLOC in 2024 to 19,726 SLOC by September 2026~\cite{rfl_empirical_2024}.


The increasing complexity of this code is exemplified by Android Binder: Google's Rust Binder driver entered mainline Linux in version 6.18, with ongoing work toward replacing the legacy C implementation~\cite{rust-binder-rfc,rust-binder-linux-6-18,binder-c-removal}. 
This rapid evolution—from basic language support to complex, highly concurrent drivers—makes understanding the concurrency correctness of Rust kernel code increasingly important.

\begin{figure}[!t]
    \centering
    \includegraphics[width=0.8\textwidth]{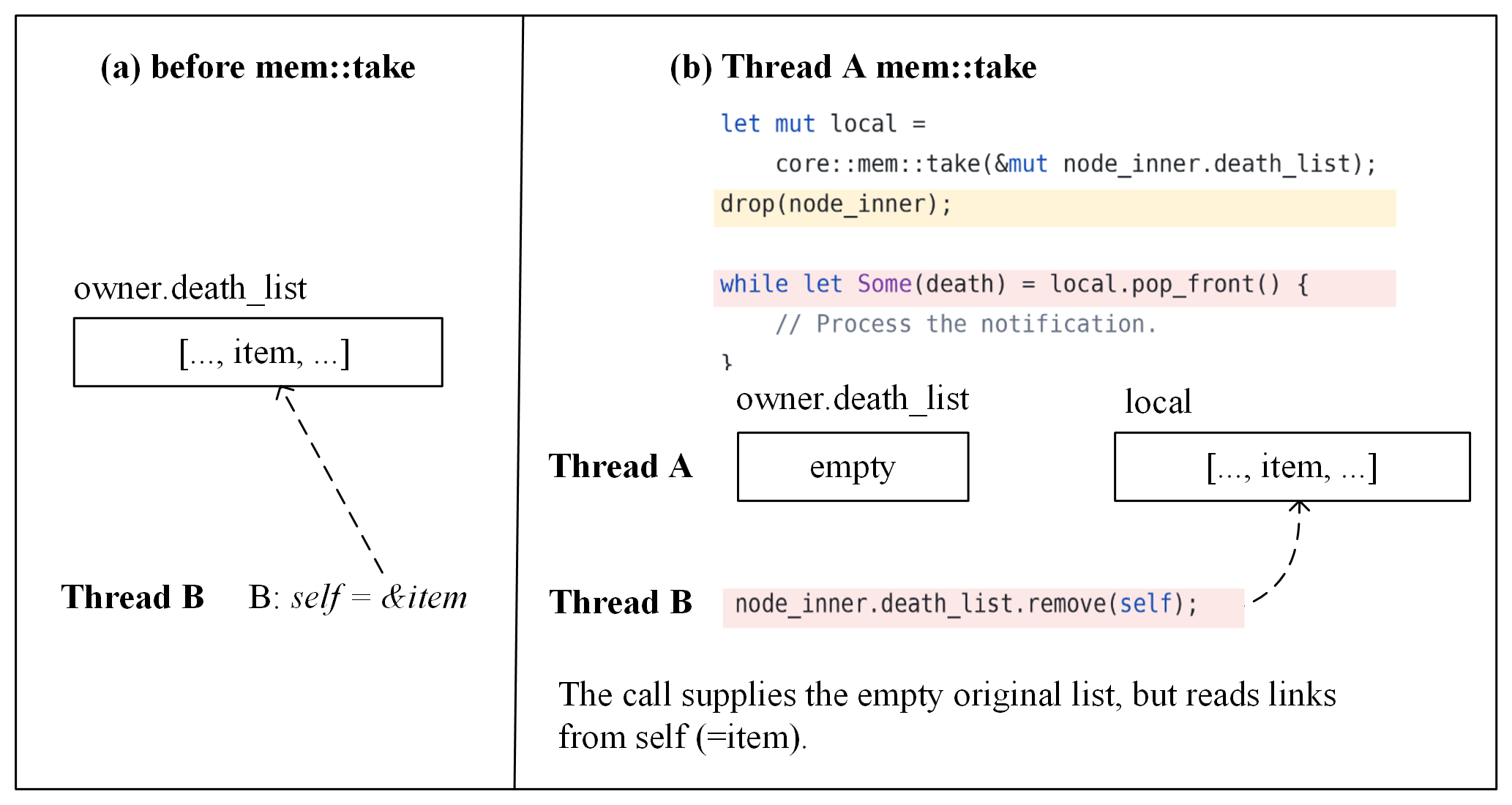}
    \caption{\textbf{CVE-2025-68260 in the Linux Rust Binder driver.} A accesses links via local. B accesses links via self. The owner lock does not guard the local list.}
    \label{fig:cve}
\end{figure}

However, Rust's memory- and thread-safety guarantees do not imply that Linux drivers written in Rust are free of concurrency bugs. In 2025, a severe vulnerability (CVE-2025-68260) was reported in the Rust Binder driver~\cite{cve_2025_68260}, as illustrated in Figure~\ref{fig:cve}. The vulnerability arose from an invalid assumption about list ownership and lock protection. Concurrent operations are allowed and corrupt the notification list. This example highlights an important limitation of Rust's safety guarantees: they do not ensure the \emph{logical correctness of concurrent operations}. A driver may be free of conventional data races yet still behave incorrectly when individually valid operations interleave in a way that violates a higher-level protocol or state invariant.

We call such a failure a \emph{logical race}: an execution in which the ordering or interleaving of concurrent operations invalidates a logical assumption on which one or more operations depend. As increasingly complex Linux components are implemented in Rust, detecting such bugs becomes increasingly important because they fall outside the class of concurrency errors that Rust's type system is designed to prevent.

In this case, Binder maintains a list of death notifications associated with each node. During node destruction, Node::release moves these notifications to a temporary local list and releases the node lock before processing them. A concurrent operation may nevertheless attempt to remove a notification under the assumption that it is either linked to its owner's list or not linked to any list. Moving the notification to the temporary list violates this assumption: the notification remains linked, but is no longer protected by the lock expected by the concurrent operation. The two operations may therefore manipulate the same notification under incompatible assumptions about its state, eventually corrupting the underlying list. This example illustrates a broader challenge: logical races arise from violations of program-specific invariants that are often implicit in the code rather than enforced by Rust's type system.


Detecting logical races of this kind is non-trivial for several reasons:

\begin{itemize}
\item \noindent\textbf{Challenge 1} 
Concurrency testing requires an oracle that distinguishes an incorrect interleaving from a valid concurrent execution. One approach is to manually specify the intended behavior. E.g., CONVEROS~\cite{converos_2025} verifies 12 concurrency modules in the Rust-based OS Asterinas~\cite{asterinas_2025} using PlusCal specifications derived from its implementation. Such specifications can express rich safety properties, but constructing and maintaining them requires substantial effort and expertise. Kernel concurrency fuzzers can expose failures through controlled schedule exploration. E.g., SECT~\cite{xu2026sect} injects scheduling points through LLVM instrumentation and uses a \texttt{sched\_ext}-based eBPF scheduler to explore kernel thread interleavings. Such testing, however, still requires an observable failure condition or other oracle to determine whether a particular execution is incorrect. For logical races, the challenge is therefore not only to expose unusual interleavings, but also to determine automatically whether their behavior is semantically invalid.

\item \noindent\textbf{Challenge 2} 
The number of possible executions grows rapidly with the number of threads and scheduling points. Controlled concurrency testing addresses this problem by repeatedly executing a harness under different controlled schedules. Loom~\cite{loom} systematically explores valid execution permutations while applying state-reduction techniques, whereas Shuttle~\cite{shuttle} emphasizes randomized schedule exploration together with deterministic replay of failures. Nevertheless, their effectiveness depends critically on which operations a harness exercises and where scheduling choices are introduced~\cite{pct}. Exploring every possible scheduling point is prohibitively expensive for realistic driver code; effective testing must instead identify the interleavings most likely to expose violations of logical ordering constraints.

\item \noindent\textbf{Challenge 3} 
Linux drivers depend heavily on kernel interfaces and execution contexts that are difficult to reproduce in a small, deterministic userspace test. A Rust driver does not execute in isolation: it invokes kernel interfaces implemented in both Rust and C and interacts with kernel objects, callbacks, synchronization primitives, and lifecycle management. Testing a small piece of driver logic inside a complete Linux kernel requires executing a large system in which unrelated kernel activity makes fine-grained schedule control difficult. RustMC~\cite{pearce2025rustmc} operates over LLVM IR to model-check concurrent Rust programs, including programs with C/C++ dependencies. However, controlled concurrency testing of Linux Rust drivers still requires a lightweight execution environment that faithfully reproduces the kernel dependencies and synchronization behavior relevant to the driver code under test.

\end{itemize}

To address these challenges, we propose XRust, a technique for detecting logical races in Linux Rust drivers. At the core of XRust is a new relation, \textit{Logical Happens-Before (LHB)}, which captures semantic ordering requirements between operations on shared state. Conventional happens-before characterizes ordering induced by program order and synchronization and is commonly used to reason about conflicting memory accesses. 
LHB preserves thread-local program order while omitting the mutex release-acquire ordering.
Two operations may individually use synchronization correctly and contain no conventional data race, yet an interleaving of their constituent steps may still produce behavior that cannot be explained by any valid serial execution. XRust uses this distinction to identify and test precisely such interleavings.

First, XRust automatically constructs an executable oracle for logical races. For a pair of operations on a shared state, XRust executes the operations sequentially in both possible orders from equivalent initial states and records their observable outcomes. These serial executions define the reference behaviors for the operation pair. XRust then executes the same operations concurrently under controlled schedules. If a concurrent execution produces an observable state or outcome that is inconsistent with both serial orders, XRust reports a potential logical race. This serial-equivalence oracle avoids requiring developers to manually encode a separate invariant for every operation pair while still capturing violations of their expected atomic behavior.

Second, XRust uses LHB to focus exploration on promising interleavings. Exhaustively inserting scheduling decisions at every memory access would create an impractically large search space. XRust instead performs runtime analysis of shared-state accesses and synchronization events to identify points at which the relative ordering of the two operations may affect their logical behavior. These observations are used to prioritize scheduling points around accesses to shared state and around synchronization boundaries that may separate logically related steps. XRust then guides controlled concurrency testing toward these candidate interleavings rather than exploring the entire schedule space uniformly.

Third, XRust uses a lightweight userspace abstraction to execute Linux Rust driver code under controlled schedules. Rather than running the driver inside a full Linux kernel, XRust emulates only the kernel interfaces exercised by a target harness, replacing them with lightweight userspace implementations that preserve the state, synchronization, and callback behavior relevant to concurrency testing. These implementations are constructed with assistance from Claude Code (\texttt{claude-opus-5}) and checked through compilation and manual compatibility review. The original Rust driver logic remains unchanged. Within this environment, XRust uses Shuttle to control thread scheduling and record executions that trigger oracle violations, enabling deterministic replay of reported logical races.

XRust performs its shared-state and synchronization analysis dynamically on compiled harnesses. It extends ThreadSanitizer~\cite{threadsanitizer} to observe memory accesses and synchronization relationships at runtime, avoiding the need for a whole-program static model or source-level rewriting of synchronization primitives. Combined with the lightweight userspace abstractions described above, this design allows XRust to analyze executions of real Rust driver code while retaining precise control over thread scheduling.


In summary, this paper makes the following contributions:

\begin{enumerate}
    \item We introduce Logical Happens-Before (LHB), a formulation for reasoning about semantic ordering requirements between concurrent operations. Such exposed synchronized access conflicts may be individually data-race-free yet collectively violate higher-level state invariants.

    \item We present XRust, an automated technique for detecting logical races in Linux Rust drivers. XRust constructs operation-pair harnesses and serial-equivalence oracles, guides schedule exploration toward interleavings likely to violate logical ordering constraints, and supports deterministic replay without requiring manually specified invariants.

    \item We develop a lightweight userspace approach for executing Linux Rust driver code under controlled schedules. XRust extends ThreadSanitizer to observe shared-memory accesses and synchronization at runtime and provides lightweight abstractions for the kernel interfaces required by each harness.

    \item We evaluate XRust on ten Rust drivers in Linux kernel 7.3-rc2, covering diverse synchronization mechanisms, including spinlocks, mutexes, and file locks. Across these drivers, XRust enumerates 1,735 pairs of operations that may access the same object and constructs executable harnesses for 1,654 of them. XRust discovers two previously unknown logical races in \texttt{rnull} and \texttt{Binder}. The fix for the \texttt{rnull} logical race has been accepted and applied by the Linux block subsystem maintainer.
\end{enumerate}

\section{Background}

\subsection{Logical Races}
\label{sec:logical-races}

Concurrent executions induce a partial order over program events known as the \emph{happens-before} (HB) relation. HB includes both program order within a thread and synchronization order across threads. For example, releasing a mutex and subsequently acquiring the same mutex creates an HB relation between the corresponding critical sections.

\noindent\textbf{Data race.}
A conventional data race occurs when two threads access the same memory location, at least one access is a write, and the accesses are not ordered by synchronization. ThreadSanitizer and related detectors identify such unordered memory conflicts~\cite{threadsanitizer}. Rust prevents most data races in safe code by restricting mutable aliasing and requiring shared mutable state to be accessed through synchronization abstractions.

\noindent\textbf{Logical race.}
Data-race freedom, however, does not imply concurrency correctness. Empirical studies have shown that many real-world concurrency failures arise from atomicity and ordering violations rather than individual unsynchronized memory accesses~\cite{lu2008learning, pearce2025rustmc}. 
We use the term \emph{logical race} for an interleaving-dependent violation of a program-level invariant even though the underlying memory accesses are individually synchronized. Such a failure can arise when one operation observes protected state, releases the lock, and later continues based on that observation after another thread has legally changed the state.

\begin{figure}[!htbp]
    \centering
    \includegraphics[width=0.75\textwidth]{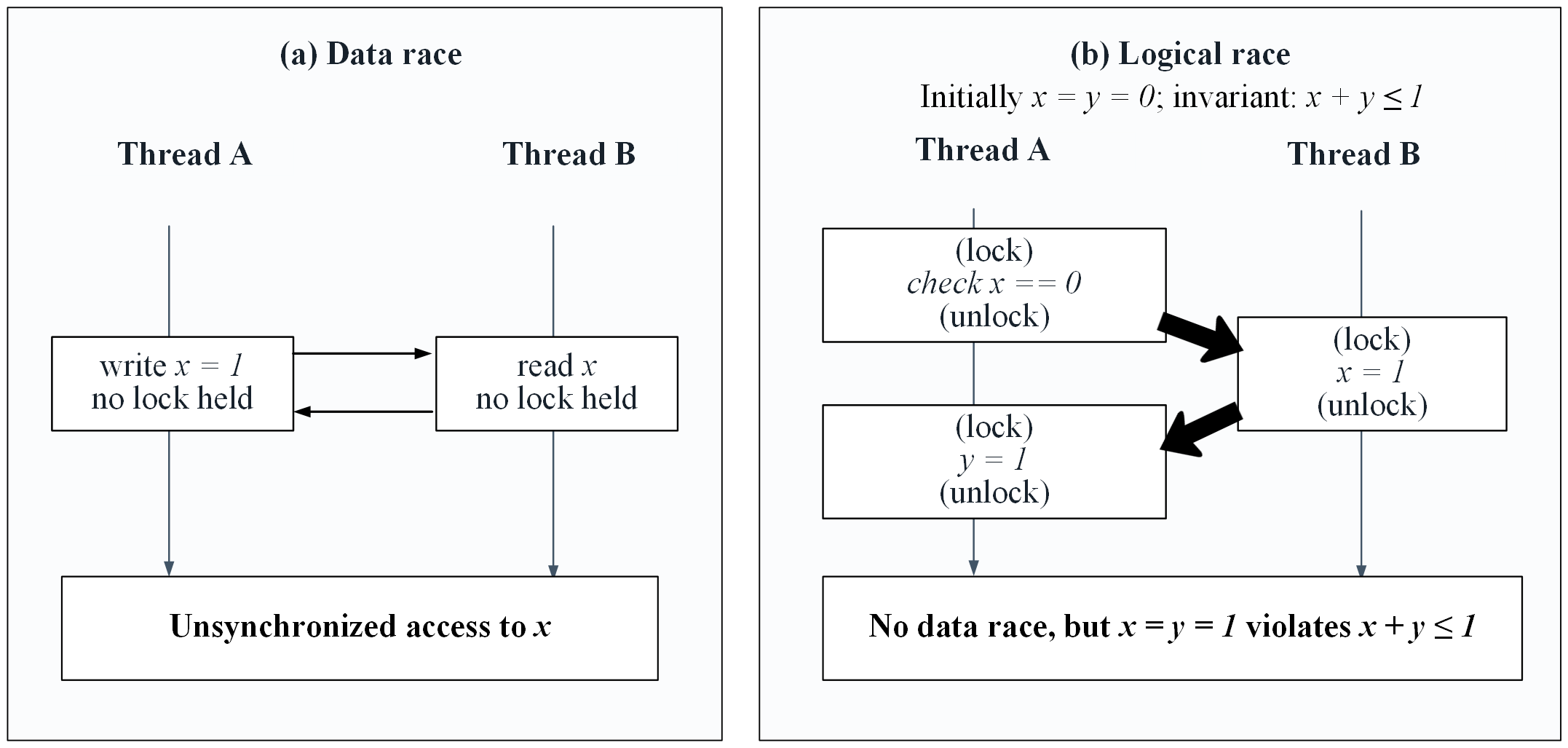}
    \caption{Data race versus logical race.}
    \label{fig:data-logical-race}
\end{figure}

Figure~\ref{fig:data-logical-race} illustrates the distinction.
In Figure~\ref{fig:data-logical-race}(a), the two threads access \texttt{x} without synchronization, producing a conventional data race. In Figure~\ref{fig:data-logical-race}(b), every access is protected by a lock, but Thread~A's check and later update are separated by a critical section executed by Thread~B. The resulting state violates $x+y\leq1$ despite containing no conventional data race.


\subsection{Rust Support in the Linux Kernel}

The Linux kernel integrates Rust as an additional implementation language alongside its predominantly C code base. A 2024 empirical study of RFL evaluated its early safety, performance, and development costs during an early stage of Rust driver development~\cite{rfl_empirical_2024}. Since then, Rust driver code in Linux kernel has grown into a meaningful subject for study. In a Linux 6.8 snapshot preceding that study, 'drivers/' contained just one Rust source file with 95 source lines of code (SLOC). As of September 2026, the Linux snapshot analyzed in this work contains 50,703 SLOC of non-vendored Rust code, including the 19,726 SLOC under 'drivers/', including implementations in the Android and GPU subsystems. Rust is increasingly moving beyond experimental kernel modules and being adopted in production-quality drivers. In several subsystems, new Rust implementations are beginning to complement or replace existing C drivers, demonstrating a gradual transition of security-critical kernel functionality from C to memory-safe Rust~\cite{binder-c-removal}.

\begin{figure}[!htbp]
    \centering
    \includegraphics[width=0.7\textwidth]{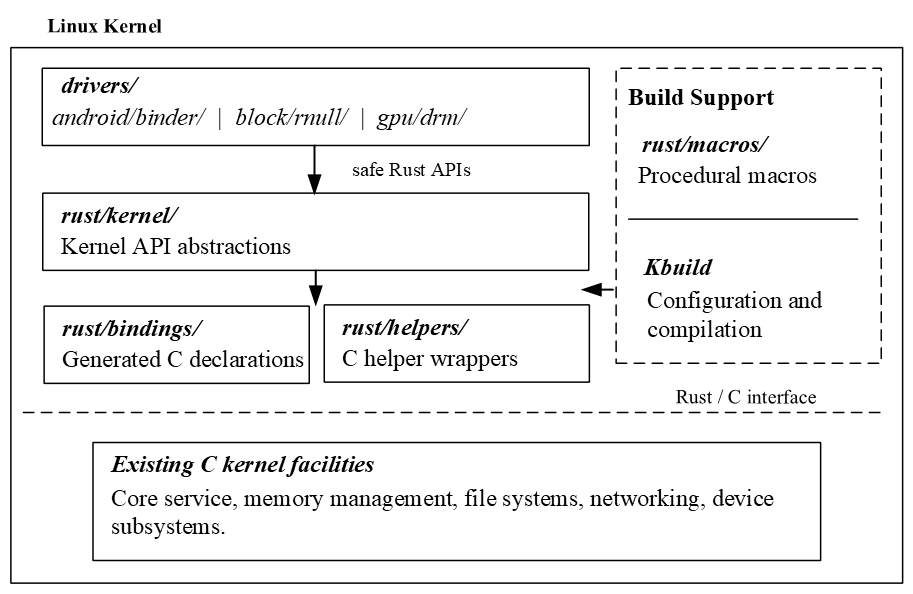}
    \caption{\textbf{Organization of Rust code in Linux kernel.}
Rust drivers under \texttt{drivers/} use kernel abstractions under \texttt{rust/}, which connect to existing C kernel facilities through bindings and helpers. Procedural macros and build integration support the Rust components.}
    \label{fig:struct}
\end{figure}

As shown in Figure~\ref{fig:struct}, in-kernel Rust code consists primarily of two parts: the Rust support infrastructure under \texttt{rust/} and selected Rust drivers under \texttt{drivers/}. The support infrastructure includes safe kernel abstractions, generated C bindings, helper functions, procedural macros, and build-system integration. Rust drivers use these abstractions to access existing kernel facilities, most of which remain implemented in C. 

\subsection{ThreadSanitizer Runtime Patch}
\label{sec:tsan-patch}

\noindent\textbf{ThreadSanitizer.}
ThreadSanitizer (TSan) is a dynamic data-race detector based on happens-before analysis. During compilation, the compiler inserts callbacks around memory accesses and synchronization operations. At runtime, TSan maintains metadata for each monitored memory location, a logical clock for each thread, and synchronization clocks for objects such as mutexes. The clock transfers create a release-acquire happens-before edge~\cite{lamport1978time, flanagan2009fasttrack}. TSan reports two accesses \(e_1\) and \(e_2\) when they originate from different threads, access the same memory location, with at least one write, and are not ordered by happens-before.

\noindent\textbf{Binary-wide applicability.}
Implementing the analysis in the TSan runtime avoids source-level rewriting of individual drivers and synchronization primitives. Any executable or library compiled with TSan instrumentation and linked against the patched runtime automatically receives transformation before analysis. The analysis therefore operates across the complete instrumented binary rather than only the source files explicitly examined by a source-level transformation. This binary-level deployment is substantially easier to apply across a heterogeneous driver corpus: the runtime is patched once, while each target only needs to be rebuilt with standard TSan instrumentation and linked against that runtime.

\section{Problem Statement}
\label{sec:problem}

Rust prevents conventional data races in safe code by disallowing unsynchronized concurrent access to mutable memory. This guarantee applies to individual accesses rather than to program-level operations spanning multiple critical sections. A thread may observe protected state, release the corresponding lock, and later continue based on that observation. During the interval, another thread may legally acquire the lock and invalidate the first thread's decision.

\noindent\textbf{Events and conflicting accesses.}
We model an execution as a schedule $\sigma$ containing memory and synchronization events. An event $e$ records its executing thread $\mathsf{tid}(e)$, memory location $\mathsf{loc}(e)$, access type $\mathsf{type}(e)\in\{\mathsf{R},\mathsf{W}\}$, and the set of locks held at the access, $\mathsf{Held}(e)$. Two events conflict when they originate from different threads, access the same location, and at least one is a write:

\begin{equation}
\begin{aligned}
\mathsf{Conflict}(e_1,e_2)
\triangleq{}&
\mathsf{tid}(e_1)\neq\mathsf{tid}(e_2)
\land
\mathsf{loc}(e_1)=\mathsf{loc}(e_2) \\
&{}\land
\bigl(
\mathsf{type}(e_1)=\mathsf{W}
\lor
\mathsf{type}(e_2)=\mathsf{W}
\bigr).
\end{aligned}
\label{eq:conflict}
\end{equation}

A conventional data race is a conflicting pair that is unordered by happens-before:

\begin{equation}
\mathsf{DataRace}(e_1,e_2)
\triangleq
\mathsf{Conflict}(e_1,e_2)
\land
\neg(e_1 \prec_{\mathrm{HB}} e_2)
\land
\neg(e_2 \prec_{\mathrm{HB}} e_1).
\label{eq:data-race}
\end{equation}

For an access event $e$, let $\mathsf{Held}(e)$ denote the locks held by its thread at that point. Two accesses share synchronization context when

\begin{equation}
\mathsf{CommonLock}(e_1,e_2)
\triangleq
\mathsf{Held}(e_1)\cap\mathsf{Held}(e_2)
\neq\varnothing.
\label{eq:common-lock}
\end{equation}


\noindent\textbf{Observable detection criterion.}
Our target is a data-race-free, cross-critical-section atomicity violation in which two individually synchronized driver operations interleave to produce an observable outcome.
Because Linux driver code generally provides no explicit specification of its operation-level invariant, we use serial equivalence as an operational criterion. For driver operations $f$ and $g$, let $O(f;g)$ and $O(g;f)$ denote the observable outcomes of the two serial orders, and let $O(f\parallel_{\sigma}g)$ denote their outcome under concurrent schedule $\sigma$. We define

\begin{equation}
\begin{aligned}
\mathsf{NonSerializable}(f,g,\sigma)
\triangleq{}&
O(f\parallel_{\sigma}g)\neq O(f;g) \\
&{}\land
O(f\parallel_{\sigma}g)\neq O(g;f).
\end{aligned}
\label{eq:nonserializable}
\end{equation}

A non-serializable outcome is not a complete specification of driver correctness. It is an executable criterion for the class of failures targeted in this work: interleavings that tear an operation across multiple critical sections and produce an observable state that cannot arise from either atomic ordering. XRust therefore uses such outcomes as candidate atomicity violations and validates the reported cases against the corresponding driver semantics.



\section{XRust Design}

\begin{figure*}[t]
    \centering
    \includegraphics[width=0.9\textwidth]{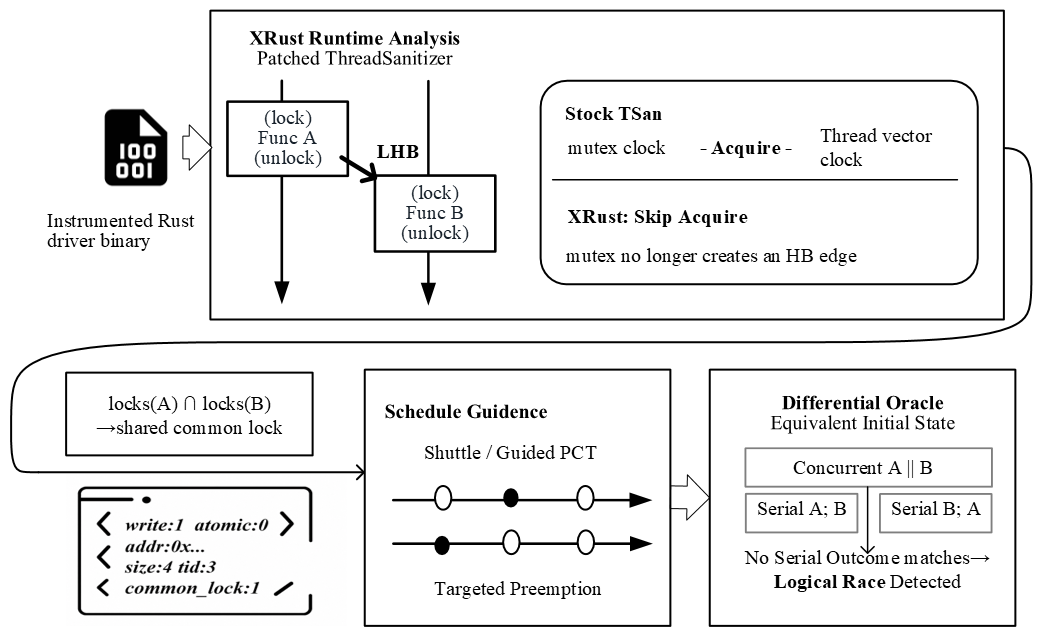}
   \caption{\textbf{XRust workflow.} Given an instrumented Rust driver binary, XRust relaxes selected happens-before edges to expose access pairs hidden from conventional race detection. These pairs are candidates for violations of logical happens-before (LHB), rather than confirmed races. XRust uses them to guide Shuttle toward relevant interleavings. The differential oracle reports a non-serializable execution when the concurrent outcome matches neither serial order from an equivalent initial state.}
    \label{fig:overview}
\end{figure*}

Figure~\ref{fig:overview} presents the overall workflow of XRust. XRust first identifies operations on shared driver state and automatically constructs concurrent harnesses together with serial-differential oracles. The oracle establishes the expected behavior of an operation pair by executing the two operations in both serial orders from equivalent initial states.

XRust then performs Logical Happens-Before (LHB) analysis to identify interleavings that may violate these serial behaviors. It selectively relaxes lock-induced happens-before relationships during analysis to expose cross-thread accesses that are individually synchronized but may participate in a logical race. For each exposed access pair, XRust records the locks held at the access sites and uses their common lock to associate the accesses with the shared protected state and corresponding critical sections.

Finally, XRust translates these candidate accesses into targeted scheduling points. A controlled scheduler prioritizes interleavings around these points while preserving the synchronization semantics of the original execution. Each concurrent outcome is checked against the serial-differential oracle; an outcome inconsistent with both serial orders is reported as a potential logical race and can be deterministically replayed.

\subsection{Harness and Oracle Generation}
\label{sec:harness-generation}

The original driver code contains neither concurrency test harnesses nor oracles for identifying invariant violations, making logical races difficult to trigger and confirm reliably. We therefore construct concurrency harnesses by grouping driver operations according to the shared state they access and materializing callable operation pairs in a common test environment.

For each shared object \(s\), let \(F_s\) denote the set of functions that acquire the lock protecting \(s\). We generate function pairs only within \(F_s\). This object-based grouping avoids pairing functions merely because they occur in the same source file.

\paragraph{Shared-state grouping.}
Each lock-protected field is identified by its enclosing type and field path.
An operation summary records the protected fields accessed by each function, and an inverse mapping associates each field with its accessing functions.

For every function \(f \in F_s\), we generate a self-pair harness \(f \parallel f\), which executes the same function concurrently on two threads. For every unordered pair of distinct functions \(\{f,g\} \subseteq F_s\), we generate a harness \(f \parallel g\). 

For rnull, the shared object is \texttt{DeviceConfig.data}, protected by a \texttt{Mutex<DeviceConfigInner>}. The corresponding function set contains the \texttt{store} and \texttt{show} operations for each of five configuration attributes:

\[
F_{\texttt{DeviceConfig.data}} =
\{
\texttt{power},
\texttt{blocksize},
\texttt{rotational},
\texttt{size},
\texttt{irqmode}
\}
\times
\{
\texttt{store},
\texttt{show}
\}.
\]

Thus, \(|F_{\texttt{DeviceConfig.data}}|=10\). Harness generation produces 10 self-pair harnesses and

\[
\binom{10}{2}=45
\]

distinct-function harnesses, for a total of 55 two-thread harnesses.

\paragraph{Harness initialization.}
Harness initialization uses the driver's default configuration or follows initialization sequences present in the original source code.
When an operation requires a particular object state, the harness performs the necessary setup operations before starting the concurrent calls.

\subsubsection{Observable Serializability Oracle}
\label{sec:auto-oracle}

For operation pairs whose relevant state can be observed, XRust uses the criterion in Equation~\ref{eq:nonserializable} as an executable oracle. Rather than requiring a separate hand-written invariant for each such pair, the harness compares one concurrent execution against the two possible serial executions of the same operations.

The criterion is intentionally operational rather than complete.
A non-serializable outcome exposes the class of atomicity violations targeted in this work: executions in which an operation is torn across multiple critical sections and produces a third observable state that cannot arise from either serial ordering.

The harness creates three independent instances initialized by the same
setup procedure. The resulting test has the following conceptual form:

\begin{quote}
\ttfamily\footnotesize
run\_pair(f, g):\\
\hspace*{1em}obj\_fg   = setup()\\
\hspace*{1em}obj\_gf   = setup()\\
\hspace*{1em}obj\_conc = setup()
\end{quote}

\vspace{-0.6em}
\begin{center}
\footnotesize
\setlength{\tabcolsep}{3pt}
\renewcommand{\arraystretch}{1.05}
\begin{tabularx}{\columnwidth}{
@{}>{\ttfamily\raggedright\arraybackslash}X
>{\ttfamily\raggedright\arraybackslash}X
>{\ttfamily\raggedright\arraybackslash}X@{}}
\textbf{Serial $f;g$} &
\textbf{Serial $g;f$} &
\textbf{Concurrent} \\
\hline
f(obj\_fg) &
g(obj\_gf) &
thread A: f(obj\_conc) \\

g(obj\_fg) &
f(obj\_gf) &
thread B: g(obj\_conc) \\

S\_fg = snapshot(obj\_fg) &
S\_gf = snapshot(obj\_gf) &
join A, B \\

&
&
S\_conc = snapshot(obj\_conc)
\end{tabularx}
\end{center}

\vspace{-0.5em}
\begin{quote}
\ttfamily\footnotesize
if S\_conc != S\_fg and S\_conc != S\_gf:\\
\hspace*{1em}report non-serializable execution
\end{quote}

The three instances are initialized through the same driver-specific setup routine. A snapshot records the observable state relevant to the tested shared object together with operation return values or errors when these affect the observable result. The comparison therefore concerns logical state rather than object identity.

The differential comparison also serves as a filter over the conservative XRust candidates. Candidate detection reasons primarily about conflicting accesses and happens-before ordering, while some interactions are constrained by higher-level mechanisms, such as reference counting or object-lifetime conditions. In such cases, XRust may retain the access pair as a candidate, but the serializability oracle does not report a violation as long as the concurrent snapshot matches either $S_{fg}$ or $S_{gf}$.

An oracle hit is treated as a candidate atomicity violation.
We inspect the corresponding driver semantics before reporting it as a
confirmed logical race.

\subsection{LHB-Based Candidate Detection}
\label{sec:candidate-detection}

XRust uses logical happens-before (LHB) analysis to identify synchronized memory-access pairs that may participate in a logical race. Its candidate detector selectively relaxes happens-before edges in ThreadSanitizer’s vector-clock analysis, and preserves thread-local program order. By recording the locks held at each access, XRust exposes conflicting pairs that conventional ThreadSanitizer would suppress and retains their synchronization context for subsequent testing.

\subsubsection{Logical Happens-Before (LHB) Analysis}
\label{sec:lhb}

To recover candidate pairs that access shared states, XRust applies logical happens-before (LHB) analysis to ThreadSanitizer's vector clocks. 

XRust maintains a vector clock $C_t$ for each thread $t$ and a vector clock $C_L$ for each mutex $L$. The operator $\sqcup$ denotes the component-wise maximum. In conventional HB analysis, releasing a mutex publishes the thread's causal history, and a subsequent acquisition imports it:
\begin{equation}
C_L \gets C_t
\qquad\text{and}\qquad
C_t \gets C_t \sqcup C_L.
\label{eq:mutex-hb}
\end{equation}
Thus, if thread $t_1$ releases $L$ before thread $t_2$ acquires it, accesses in their respective critical sections are HB-ordered. ThreadSanitizer does not report a conflicting pair protected by this synchronization.

\textbf{logical happens-before (LHB)} We define XRust's logical happens-before (LHB) relation by removing mutex release-acquire edges from the conventional HB relation.
Let $G_{\mathrm{HB}}$ be the conventional happens-before graph, and let $E_{\mathrm{mutex}}$ be the mutex release--acquire edges relaxed by XRust. We call an access pair $(a,b)$ an LHB candidate when
\begin{equation}
a \leadsto b \text{ in } G_{\mathrm{HB}}
\quad\land\quad
a \not\leadsto b \text{ in }
G_{\mathrm{HB}}\setminus E_{\mathrm{mutex}},
\label{eq:lhb-candidate}
\end{equation}
where $\leadsto$ denotes reachability.
LHB is an analysis relation for candidate discovery, and does not establish a logical race.

XRust implements this relation by omitting the mutex-clock join at mutex acquisitions in its candidate-detection analysis:
\begin{equation}
\widetilde{C}_t \gets \widetilde{C}_t
\qquad\text{rather than}\qquad
\widetilde{C}_t \gets \widetilde{C}_t \sqcup C_L.
\label{eq:lhb-acquire}
\end{equation}
Here, $\widetilde{C}_t$ is the relaxed clock. XRust still advances each thread's clock, preserving program order within the thread. The driver continues to acquire and release the mutex, so the critical sections remain mutually exclusive at runtime.

In parallel with the relaxed vector-clock analysis, XRust records each mutex acquisition in the thread's runtime lockset.
The relaxed clock makes a previously hidden cross-thread access pair visible, while the locksets show whether the accesses share a protecting mutex.

\subsubsection{Read-Write Conflicts}
\label{sec:read-write-conflicts}

XRust next filters LHB-exposed accesses by access type and memory location. We are particularly targeting interactions with at least one write:
one operation reads protected state and may make a decision based on that value, while another operation modifies the same state. XRust retains conflicting pairs as defined in Equation~\ref{eq:conflict}. In the check-then-act patterns central to our study, these conflicts commonly appear as one operation reading protected state and another operation modifying that state.

\subsubsection{Common-Lock Filtering}
\label{sec:common-lock}

LHB analysis exposes conflicting accesses whose conventional HB ordering depends on mutex synchronization. To retain the synchronization context that produced this ordering, XRust records the lockset held by each thread at every candidate access.

Using the definition of $\mathsf{CommonLock}$ from Equation~\ref{eq:common-lock}, XRust retains an LHB-exposed conflict only when the two accesses were observed while holding at least one common lock. This distinguishes the targeted pattern from arbitrary conflicts that become unordered for unrelated reasons.

XRust therefore retains an access pair when

\begin{equation}
\begin{aligned}
\mathsf{Candidate}(e_1,e_2)
\triangleq
\mathsf{Conflict}(e_1,e_2)
\land
\{e_1,e_2\}\in\mathcal{C}_{\mathrm{LHB}}
\land
\mathsf{CommonLock}(e_1,e_2).
\end{aligned}
\label{eq:xrust-candidate}
\end{equation}

$\mathcal{C}_{\mathrm{LHB}}$ denotes the set of unordered access pairs satisfying the LHB candidate condition in Equation~\ref{eq:lhb-candidate}.

For nested critical sections, the recorded intersection may contain more than one lock. Each candidate record contains the access types, source locations, runtime frames, and common-lock identities. These records are subsequently mapped to the driver operations used for schedule guidance.

\subsection{Candidate-Guided Schedule Exploration}
\label{sec:candidate-guided-scheduling}

Candidate detection identifies synchronized memory accesses that may participate in a logical race, but it does not establish that a violating interleaving is feasible. XRust therefore uses the candidate access sites to guide controlled concurrency testing with Shuttle~\cite{shuttle}. Shuttle executes the same harness repeatedly under a user-space scheduler by replacing standard threading, synchronization, and atomic primitives with instrumented equivalents. This exposes scheduling decisions to the runtime and allows failing executions to be replayed deterministically ~\cite{godefroid2005dpor,musuvathi2008chess,sen2008race}.

Shuttle supports probabilistic concurrency testing (PCT), which explores executions by assigning priorities to runnable threads and changing those priorities at a bounded number of scheduling points~\cite{pct}. In unguided PCT, these priority changes may be
selected from all eligible scheduling operations. A realistic driver harness can contain many such operations, most unrelated to the shared state interaction identified by XRust. Consequently, the probability of choosing the few decisions needed to expose a candidate interaction decreases as unrelated scheduling points accumulate.

\paragraph{Scheduling probes.}
XRust inserts scheduling probes at all eligible scheduling sites throughout the harness and the driver source code.
These sites are where PCT scheduler chooses to select other runnable threads.
For each conflicting access, XRust walks up the recorded call stack to identify the corresponding call sites in the driver and harness code, then matches their source locations to the inserted probes.

\paragraph{Targeted Preemption}
\label{sec:targeted-preemption}

XRust changes where PCT concentrates its scheduling decisions. For each candidate pair, XRust records the source locations of the two conflicting accesses. The instrumented harness contains dormant scheduling probes at eligible driver locations; only probes matching the candidate sites are activated for the corresponding test. Reaching an inactive probe has no scheduling effect, whereas an active probe allows Shuttle to suspend the current thread and schedule another runnable thread.

For example, suppose Thread~A reads protected state and later acts on that observation, while Thread~B contains the conflicting update. After Thread~A performs the candidate read, an active probe gives the scheduler an opportunity to run Thread~B before Thread~A continues. That is, The competing operation can enter its critical section after the reader releases the protecting lock. The resulting execution preserves the driver's synchronization semantics but increases the chance of exposing an interleaving in which Thread~A acts on stale state.
XRust therefore narrows the scheduling locations at which PCT concentrates its priority changes.

\section{Results}

\subsection{Linux Rust Corpus}
\label{sec:linux-rust-corpus}

We evaluate XRust on Linux 7.3-rc2~\cite{linux-v7.3-rc2}, whose source tree contains 496 Rust source files. Our driver evaluation covers the 102 Rust files in the ten driver sets listed in Table~\ref{tab:linux-rust-corpus}. Vendored dependencies, build-time crates and in-kernel standard library are excluded.

XRust first identifies driver operations that access shared mutable state through a lock or atomic object. We distinguish a \emph{unique function} from a \emph{function-object incidence}: one function may access more than one shared object and therefore participate in multiple object-specific pair sets. Across the corpus, XRust identifies 139 unique functions and 161 function--object incidences over 17 shared objects.

For each shared object, XRust enumerates unordered pairs of operations that access that object, including self-pairs. A self-pair executes the same operation concurrently in two threads; a same-file pair contains two distinct operations defined in the same file; and a cross-file pair contains operations defined in different files. This enumeration produces 1,735 \emph{planned pairs}. 

\begin{table*}[t]
    \centering
    \caption{Shared-state operations and planned function pairs in the
    Linux 7.3-rc2 Rust driver corpus. ``U/O'' denotes unique functions
    and function--object incidences, respectively.}
    \label{tab:linux-rust-corpus}
    \small
    \setlength{\tabcolsep}{5pt}
    \begin{tabularx}{\textwidth}{@{}l r X r r r r r@{}}
        \toprule
        \textbf{Driver} &
        \textbf{Files} &
        \textbf{Shared state} &
        \textbf{Funcs. U/O} &
        \textbf{Self} &
        \textbf{Same} &
        \textbf{Cross} &
        \textbf{Planned} \\
        \midrule

        Binder &
        20 &
        12 shared fields &
        116/138 &
        138 &
        1,163 &
        349 &
        1,650 \\

        rnull &
        2 &
        \texttt{DeviceConfig.data} &
        10/10 &
        10 &
        45 &
        0 &
        55 \\

        Nova &
        4 &
        DRM \texttt{File} GEM handle table &
        3/3 &
        3 &
        1 &
        2 &
        6 \\

        Tyr &
        12 &
        \texttt{Mmu.as\_manager};
        \texttt{Vm.gpuvm\_unique} &
        7/7 &
        7 &
        11 &
        0 &
        18 \\

        Nova Core &
        59 &
        \texttt{Cmdq.inner} &
        3/3 &
        3 &
        3 &
        0 &
        6 \\

        rcpufreq\_dt &
        1 & None & 0/0 & 0 & 0 & 0 & 0 \\

        drm\_panic\_qr &
        1 & None & 0/0 & 0 & 0 & 0 & 0 \\

        ax88796b &
        1 & None & 0/0 & 0 & 0 & 0 & 0 \\

        qt2025 &
        1 & None & 0/0 & 0 & 0 & 0 & 0 \\

        pwm\_th1520 &
        1 & None & 0/0 & 0 & 0 & 0 & 0 \\

        \midrule
        \textbf{Total} &
        \textbf{102} &
        \textbf{17 shared objects} &
        \textbf{139/161} &
        \textbf{161} &
        \textbf{1,223} &
        \textbf{351} &
        \textbf{1,735} \\
        \bottomrule
    \end{tabularx}
\end{table*}

Five single-file drivers produce no candidate pairs because we found no driver-owned lock or atomic state shared by concurrently callable operations. 

Pair enumeration does not imply that every planned pair becomes an independent executable harness. We distinguish the static planning space from the tests that were successfully materialized and executed.

We exclude tests that could not be executed without modifying the original driver source. In particular, we retain ordinary functions and methods defined in modules when the generated harness can directly invoke the original function. By contrast, crate-root-only logic in 'rust\_binder\_main.rs' may involve private types or private functions, or may only be reachable through kernel BinderFS callbacks. Because exercising such code requires modifying the original source or introducing additional test-specific entry points, we exclude these cases from the current evaluation. 
Overall, XRust enumerates 1,735 planned same-object pairs and obtains 1,654 completed test executions; one additional Binder harness times out.

\subsection{Confirmed Logical Races}
\label{sec:confirmed-races}

The candidate analysis and controlled executions produced two logical races that we manually confirmed against the corresponding driver state transitions. Both are data-race-free: every individual access to the affected state is protected by the intended synchronization primitive. 
The \texttt{rnull} finding has been confirmed by Linux maintainers. The corresponding patch was included in the Rust block-layer series submitted for Linux v7.4.

\subsubsection{R1: rnull Geometry Update During Power-On}

The rnull driver stores configuration state in \texttt{DeviceConfig.data}, a \texttt{Mutex<DeviceConfigInner>}. A live disk copies the configured block size when the disk is created. The intended behavior is that geometry cannot change while the disk is powered.

However, the block-size setter can perform its powered-state check and its update in two different critical sections. The following simplified excerpt preserves the relevant locking structure:

\begin{lstlisting}[language=Rust,basicstyle=\ttfamily\footnotesize]
fn store_blocksize(&self, page: &[u8]) -> Result {
    if self.data.lock().powered {
        return Err(EBUSY);
    }
    let value = parse_u32(page)?;
    self.data.lock().block_size = value;
    Ok(())
}
\end{lstlisting}

In contrast, the power operation holds the same mutex while creating the disk and setting \texttt{powered}. The harness first configures an unpowered device with \texttt{block\_size=512}, and then executes \texttt{power.store("1")} concurrently with \texttt{blocksize.store("4096")}. The following execution is feasible:

\begin{enumerate}
    \item \texttt{blocksize.store} acquires the mutex, observes
          \texttt{powered=false}, and releases the mutex;
    \item \texttt{power.store} creates the live disk using the current
          block size 512 and sets \texttt{powered=true};
    \item \texttt{blocksize.store} reacquires the mutex and changes the
          configuration to 4096.
\end{enumerate}

The final state contains a powered disk whose live block size is 512 while the stored configuration says 4096. All individual accesses are mutex protected, but the check and update in \texttt{blocksize.store} are not atomic with respect to power-on.

In the serial order \texttt{power; blocksize}, the second operation observes a powered disk and returns \texttt{EBUSY}, leaving the configuration at 512. In the opposite order, \texttt{blocksize; power}, the disk is created with block size 4096. The concurrent state $\langle\text{disk}=512,\text{config}=4096\rangle$ matches neither serial execution.

\paragraph{Fix strategy.}
The check and update must execute within a single critical section. The setter acquires \texttt{DeviceConfig.data}, checks \texttt{powered}, and retains the same mutex guard until the configuration update is complete.
This restores atomicity between geometry updates and power-on without introducing an additional lock.
The fix was accepted and applied by the Linux block subsystem maintainer.

\subsubsection{B1: Binder Allocation Resurrection After Cancellation}

The second race occurs in Binder's \texttt{Transaction} state. Each transaction stores an optional allocation behind a spinlock:

\begin{lstlisting}[language=Rust,basicstyle=\ttfamily\footnotesize]
allocation: SpinLock<Option<Allocation>>
\end{lstlisting}

\texttt{prepare\_file\_list} temporarily removes the allocation from the protected field while translating file descriptors, and restores it on success. \texttt{cancel}, meanwhile, removes and destroys the allocation:

\begin{lstlisting}[language=Rust,basicstyle=\ttfamily\footnotesize]
prepare_file_list():
    alloc = self.allocation.lock().take()
    translate_fds(alloc)              // lock not held
    *self.allocation.lock() = Some(alloc)

cancel():
    alloc = self.allocation.lock().take()
    drop(alloc)
\end{lstlisting}

Each individual access to \texttt{allocation} is synchronized. Nevertheless, removing the object, operating on it outside the lock, and later restoring it creates a window in which cancellation can occur.

Starting from \texttt{allocation=Some(a)}, XRust exposes the following schedule:

\begin{enumerate}
    \item \texttt{prepare\_file\_list} takes \texttt{a}, leaving the
          protected field as \texttt{None};
    \item \texttt{cancel} acquires the lock, observes
          \texttt{None}, and therefore has no allocation to destroy;
    \item the translation in \texttt{prepare\_file\_list} succeeds;
    \item \texttt{prepare\_file\_list} reacquires the lock and stores
          \texttt{Some(a)} back into the transaction.
\end{enumerate}

The transaction therefore contains a live allocation after \texttt{cancel} has completed---an allocation resurrection.

This outcome cannot occur in either serial order. For \texttt{prepare; cancel}, preparation restores the allocation before cancellation removes it, so the final state is \texttt{None}. For \texttt{cancel; prepare}, cancellation removes the allocation first and preparation immediately observes \texttt{None}, again leaving the final state as \texttt{None}. The concurrent execution instead terminates with \texttt{allocation=Some(a)}. 

\paragraph{Fix strategy.}
A possible fix is to distinguish cancellation from the temporary absence of an allocation. A cancellation flag can be protected by the same spinlock as the allocation field.
The \texttt{cancel} operation sets this flag even when the allocation is temporarily held by \texttt{prepare\_file\_list}.
Before restoring the allocation, preparation checks the flag under the same lock and restores it only if cancellation has not occurred; otherwise, it releases the lock and drops the allocation.

\subsection{Evaluation of XRust Race Detection}

As shown in Table~\ref{tab:guided-vs-unguided}, both schedulers eventually detected the two logical races across all 20 seeds. XRust-guided Shuttle exposed each race in the first iteration for every seed, whereas PCT required up to 14 iterations for rnull and 27 iterations for Binder. This result indicates that XRust guidance improves detection speed by concentrating scheduling decisions near candidate accesses.
\begin{table}[!htbp]
    \centering
    \caption{First-failing iterations for XRust-Shuttle and PCT-Shuttle. Both methods detected each race for all 20
    shared seeds (\texttt{ITERS=50}, \texttt{DEPTH=3}).}
    \label{tab:guided-vs-unguided}
    \small
    \setlength{\tabcolsep}{4pt}
    \begin{tabular}{llrrrr}
        \toprule
        \textbf{Driver} &
        \textbf{Scheduler} &
        \textbf{Min} &
        \textbf{Med.} &
        \textbf{Mean} &
        \textbf{Max} \\
        \midrule

        \multirow{2}{*}{R1}
        & XRust-Shuttle
        & \textbf{1}
        & \textbf{1}
        & \textbf{1.00}
        & \textbf{1} \\
        & PCT-Shuttle
        & 2
        & 4
        & 5.35
        & 14 \\
        \midrule

        \multirow{2}{*}{B1}
        & XRust-Shuttle
        & \textbf{1}
        & \textbf{1}
        & \textbf{1.00}
        & \textbf{1} \\
        & PCT-Shuttle
        & 2
        & 4
        & 6.80
        & 27 \\
        \bottomrule
    \end{tabular}

    \vspace{2pt}
    \begin{minipage}{0.98\columnwidth}
        \footnotesize
        \textbf{R1:} rnull,
        \texttt{power\_store} $\parallel$ \texttt{blocksize\_store}.
        \quad
        \textbf{B1:} Binder,
        \texttt{prepare\_file\_list} $\parallel$ \texttt{cancel}.
    \end{minipage}
\end{table}

\subsection{Scaling to Realistic Transaction Workloads}
\label{sec:realistic-harness}

Our experiment used automatically generated two-function harnesses, which provide a controlled setting for comparing unguided PCT with XRust-guided scheduling. To evaluate whether the advantage persists in a more realistic workload, we constructed more complex harnesses from methods implemented by Binder's \texttt{Transaction} object. These harnesses surround the same candidate pair with actual driver operations that read or modify the shared transaction state.

For each harness, two threads operate on the same \texttt{Transaction} instance and execute an identical prefix of real \texttt{transaction.rs} methods. Only their final operations differ: Thread~A invokes \texttt{prepare\_file\_list}, while Thread~B invokes \texttt{cancel}. The split final operations preserve an observable difference between the two orderings while allowing the prefix to increase the surrounding workload.

We progressively extended the shared prefix by adding methods from the same driver implementation. Early variants contain state setters and observers such as \texttt{set\_outstanding}, \texttt{is\_oneway}, and \texttt{debug\_print}. These methods add realistic computation but few relevant scheduling decisions because the observers do not acquire the target locks. Later variants add \texttt{submit} and \texttt{drop\_outstanding\_txn}, which introduce additional synchronization and shared-state transitions. The largest harness also includes \texttt{do\_work}, which may consume the transaction allocation before the final candidate pair executes. Thus, increasing $N$ expands not only the scheduling trace but also the state space that must be explored to reach the vulnerable allocation state.

We compare Shuttle's unguided PCT scheduler against XRust-guided PCT using five random seeds, 300 iterations per seed, and PCT depth three. Table~\ref{tab:realistic-transaction-workload} reports the number of iterations required to observe the first oracle failure.
Entries report mean first-failure iterations over five seeds when a failure is observed. 0/5 within 300 indicates that none of the five seeds exposed the failure within the 300-iteration budget.

\begin{table}[!htbp]
\centering
\caption{First-failure iterations as the Binder transaction harness is extended with real driver operations.}
\label{tab:realistic-transaction-workload}
\small
\setlength{\tabcolsep}{4pt}
\begin{tabular}{c p{0.38\columnwidth} c c}
\toprule
$N$ & Shared prefix extension
    & Unguided PCT
    & XRust-guided \\
\midrule
2
& None; target pair only
& $2.8$ $(2$--$4)$
& $\mathbf{1}$ \\

4--10
& \texttt{set\_outstanding} and observers
& $4$--$7$
& $\mathbf{1}$ \\

12
& Add \texttt{submit} and \texttt{drop\_outstanding\_txn}
& $19.6$ $(2$--$38)$
& $\mathbf{1}$ \\

13
& Add \texttt{do\_work}
& $0/5$ within $300$
& $\mathbf{70.6}$ \\
\bottomrule
\end{tabular}
\end{table}

The observer-heavy prefixes have little effect on unguided PCT because they introduce few additional synchronization points. Once
\texttt{submit} and \texttt{drop\_outstanding\_txn} are included,
however, the mean first-failure iteration increases to 19.6 because PCT must select the critical ordering from a larger set of real scheduling decisions. With \texttt{do\_work}, unguided PCT fails to expose the race in all five 300-iteration runs. This method can consume the allocation before the final pair, making the vulnerable state itself rare in addition to making the required interleaving difficult to select.

XRust-guided scheduling finds the violation in the first iteration for all harnesses through $N=12$ because it concentrates preemptions at the remaining \texttt{prepare\_file\_list} and \texttt{cancel} sites. Its mean increases to 70.6 iterations at $N=13$ because scheduling guidance identifies \emph{where} to preempt but does not guarantee that execution reaches the required state. $N$ denotes the number of distinct driver methods exercised by the harness.

At this point, exposing the bug requires more than selecting the correct preemption points from a large scheduling trace. The execution must also construct the state required by the candidate interaction:

\begin{equation}
\underbrace{\mathsf{ReachVulnerableState}}_{\text{state-space search}}
\;\land\;
\underbrace{\mathsf{ChooseFailingInterleaving}}_{\text{schedule search}}.
\label{eq:state-and-schedule-search}
\end{equation}

\section{Discussion}
\label{sec:discussion}

\subsection{Executing Linux Rust Driver Logic in User Space}
\label{sec:userspace-execution}

Linux drivers cannot execute directly in a user-space concurrency-testing framework because they depend on kernel-only types, allocation routines, synchronization primitives. To make the driver logic executable under XRust and Shuttle, we construct a user-space environment containing compatibility shims and thin peer modules for the kernel dependencies imported by each target driver. We use Claude Code (\texttt{claude-opus-5}) to assist in constructing this compatibility code. The generated code is checked through compilation and manually inspected for compatibility with the tested driver operations.

The compatibility layer implements only the neighboring APIs imported by the selected driver files. For example, a shim provides the interface of a kernel mutex or reference-counted object. Similarly, thin peer modules retain only the definitions required to compile and execute the target code. This approach keeps the executable environment small while allowing the analyzed driver methods to remain source-identical to their in-kernel implementations.

Consequently, our results establish that a logical race exists in the preserved driver logic under the modeled environment. They do not prove that every generated execution is reachable through the complete kernel interface. We reduce this risk by keeping the shims thin, and preserving synchronization and shared-state operations.

\subsection{Cost Trade-offs and Scalability}
\label{sec:cost-tradeoffs}

XRust-guided scheduling reduces the number of schedules required to expose a logical race, but it introduces a fixed preprocessing cost. Its end-to-end running time can be expressed as

\begin{equation}
\begin{aligned}
T_{\mathrm{XRust}} ={}&
T_{\mathrm{detection}}
+ T_{\mathrm{instrumentation}}
+ T_{\mathrm{compilation}} \\
&{}+
T_{\mathrm{guided\ testing}}.
\end{aligned}
\label{eq:xrust-cost}
\end{equation}

By comparison, unguided PCT requires no XRust analysis and has an approximate cost of

\begin{equation}
T_{\mathrm{PCT}}
=
N_{\mathrm{iterations}}
\cdot
T_{\mathrm{iteration}},
\label{eq:pct-cost}
\end{equation}

where $N_{\mathrm{iterations}}$ is the number of schedules executed before a failure is observed or the testing budget is exhausted.

On a small harness, one Shuttle iteration is inexpensive. In our experiments, even 150 unguided PCT iterations require approximately 0.2 seconds, while XRust must additionally perform candidate detection, source instrumentation, and compilation. Pure Shuttle can therefore have lower wall-clock latency despite requiring substantially more iterations. 

The relevant crossover occurs when the accumulated cost of unguided schedule exploration exceeds XRust's fixed preprocessing cost:

\begin{equation}
N_{\mathrm{PCT}}T_{\mathrm{iteration}}
>
T_{\mathrm{detection}}
+
T_{\mathrm{instrumentation}}
+
T_{\mathrm{compilation}}
+
N_{\mathrm{guided}}T_{\mathrm{iteration}}.
\label{eq:cost-crossover}
\end{equation}

Our experiments show that the iteration count of unguided PCT increases as real synchronization and state-changing operations are added to the harness, whereas guided testing continues to find the race with fewer iterations. We expect the crossover to become more favorable to XRust for large drivers containing hundreds or thousands of scheduling points, or when the same instrumented binary is reused to test multiple candidate pairs. Thus, the current evidence primarily demonstrates a reduction in schedule-search cost; establishing an end-to-end wall-clock advantage at kernel scale requires larger and longer-running workloads.

\subsection{XRust for Migrating C Drivers}
\label{sec:xrust-for-c-drivers}

Most Linux drivers are still written in C, so we conducted a preliminary study to examine whether XRust could also be useful for future C-to-Rust driver migration.
We examined all 707 C files in \texttt{drivers/iio}, where direct device accesses must be rejected while buffered streaming is active. 
We identified 12 sites that use \texttt{iio\_buffer\_enabled()} as an unlocked rejection guard rather than the atomic \texttt{iio\_device\_claim\_direct()} interface. Since mainline Linux currently provides no Rust IIO abstraction, we recreated the relevant synchronization structure in our userspace environment and ported representative drivers from this pattern. 
The same check-then-act race remained after migration: XRust localized the conflicting accesses when the unlocked guard was preserved and stopped reporting the pair when the guard was replaced by the atomic interface. This preliminary result suggests that logical concurrency defects can survive migration to memory-safe Rust.

\begin{table}[!t]
    \centering
    \caption{Confirmed logical races among the 12
    \texttt{iio\_buffer\_enabled()} rejection guards in
    \texttt{drivers/iio}. These are the race candidates the serializability oracle retained.}
    \label{tab:iio-guards}
    \small
    \setlength{\tabcolsep}{4pt}
    \begin{tabular}{llrl}
        \toprule
        \textbf{Driver} &
        \textbf{Guarded action} &
        \textbf{XRust} &
        \textbf{Shuttle} \\
        \midrule
        \texttt{stk8312}   & mode register $\rightarrow$ standby & 2 & race \\
        \texttt{stk8ba50}  & \texttt{POWMODE} $\rightarrow$ suspend & 2 & race \\
        \texttt{ina2xx}    & \texttt{CONFIG} rewritten mid-capture & 2 & race \\
        \bottomrule
    \end{tabular}
\end{table}




\section{Related Work}
Existing tools have explored concurrent test generation, atomicity checking, and controlled schedule exploration. These techniques provide important foundations for detecting logical races, but applying them to Linux Rust drivers can still require manual construction of specifications.
Tools such as Miri~\cite{jung2026miri} can detect undefined behavior without user-written assertions. But detecting logical races with tools such as RustMC~\cite{pearce2025rustmc} and SECT~\cite{xu2026sect}, can require manual assertions or separate oracles such as KASAN.
XRust addresses this gap by connecting all components through shared-state-based harness generation, LHB-guided scheduling, and serial-outcome comparison in a lightweight userspace environment.

In this section, we walk through design choices encountered in an automatic workflow for discovering logical races in Linux Rust driver workspaces. We discuss systems described in prior work, how their designs lead to trade-offs across various quality attributes that \textsc{XRust} considers, and what \textsc{XRust} adopts or improves upon.

\paragraph{Concurrent Test Generation.}
Ballerina~\cite{ballerina} generates concurrent unit tests consisting of a sequential prefix that constructs and initializes shared objects, followed by two threads that each invoke one method. This structure limits the cost of schedule exploration while allowing the prefix to establish states needed to expose concurrency bugs. 
ConTeGe~\cite{contege} generates sequential initialization and concurrent method-call sequences on a shared object. It reports exceptions or deadlocks observed concurrently when the same failure cannot occur in any serial ordering.
These approaches establish automatic concurrent test
generation in small concurrent test scenarios.

XRust adopts the shared-object, small-harness testing strategy of these approaches. Its contribution is to organize operations around shared lock-protected state and generate both self-pair and distinct-function-pair harnesses. These harnesses are then connected to LHB-based candidate analysis and serial-outcome checking in a lightweight userspace environment.

\paragraph{Atomicity Checking Oracles.}
AVIO~\cite{avio} infers access-interleaving invariants from correct executions and reports subsequent violations of those inferred expectations.
Velodrome~\cite{velodrome} checks conflict serializability of designated atomic blocks by  constructing a transactional dependency graph and detecting cycles.
XRust, however, separates candidate identification from outcome validation.
LHB analysis identifies accesses that may participate in a logical race, but an access conflict alone is not reported as a confirmed violation. 

Sequential-reference checking approaches like Line-Up~\cite{burckhardt2010lineup} automatically check deterministic linearizability by enumerating sequential behaviors of the implementation and comparing concurrent behaviors against them, avoiding a manually written sequential specification.
Violat~\cite{violat} generates self-contained tests of observational refinement and precomputes combinations of return values admitted by the reference abstract data type.
Likewise, for each instantiated function pair, \textsc{XRust} compares the concurrent snapshot, including selected protected state, with the snapshots from both serial orderings.

XRust shares the sequential-reference principle of these approaches. For each instantiated driver-operation pair, it executes both serial orders from equivalent initial states and compares their snapshots with the concurrent outcome. The snapshots include selected lock-protected state as well as operation results. This provides an executable check for the tested pair and initial state.

\paragraph{Schedule exploration and targeted triggering.}
CTrigger~\cite{ctrigger} identifies potential  unserializable interleavings involving two accesses from one thread and an interfering access from another.
Although CTrigger ignores locks, its exploration targets access triples, whereas \textsc{XRust} utilizes lock-protected accesses to connect LHB candidates.
Lock acquisition and release boundaries are used to provide scheduling locations.

Other systems provide execution-control and exploration mechanisms that offer thorough exploration or are applicable to a specific setting or environment.
Shuttle~\cite{shuttle} and Loom~\cite{loom} support controlled exploration of Rust programs through instrumented concurrency primitives.
RustMC~\cite{pearce2025rustmc} extends the GenMC  stateless model checker to Rust at the LLVM IR level.
Fray~\cite{li2025fray} uses shadow locking to control JVM thread interleavings without replacing the underlying concurrency primitives. SECT~\cite{xu2026sect} provides fine-grained concurrency testing within the Linux kernel through eBPF-based scheduling control.
SECT explores concurrency within the Linux kernel, whereas XRust executes selected Rust driver operations in a lightweight userspace environment. This enables controlled testing of preserved driver logic, but the resulting executions remain subject to the compatibility-layer assumptions discussed in Section~\ref{sec:userspace-execution}.

These systems provide mechanisms for controlling or exploring concurrent executions. XRust builds on Shuttle as its execution backend and adds LHB-guided scheduling. Our evaluation therefore compares XRust-guided scheduling with unguided PCT within the same Shuttle environment.

\section{Conclusion}
\label{sec:conclusion}

Rust eliminates broad classes of memory-safety violations and data races, but it does not guarantee that individually synchronized operations preserve higher-level program invariants. As Rust increasingly replaces C in complex Linux drivers, detecting these logical races has become a practical systems problem. Conventional happens-before race detectors suppress the relevant accesses because they are protected by locks, while unguided concurrency testing must search a large space of scheduling decisions.

This paper presented \textsc{XRust}, a directed concurrency-analysis pipeline for finding logical races in Linux Rust drivers. XRust extends ThreadSanitizer with a relaxed happens-before analysis that exposes conflicting accesses hidden by mutex synchronization while preserving program order. It records the locks held at each access, identifies candidate pairs protected by a common lock, and translates their source locations into targeted preemption points for Shuttle-based probabilistic concurrency testing. We further automate harness and oracle generation by grouping functions around shared lock-protected objects and comparing concurrent outcomes against serial execution orders.

We evaluated XRust on the Rust driver code in Linux v7.3-rc2 and confirmed logical races in two substantially different drivers: a check-then-act violation in \texttt{rnull}, and an allocation-resurrection error in Android Binder. Experiments with increasingly complex Binder workloads show that unguided PCT becomes less likely to select the critical interleaving as real synchronization and state-changing operations are added. 

These results demonstrate that memory safety and data-race freedom do not eliminate concurrency failures from safe Rust systems code. XRust recovers interactions hidden by synchronization and uses them to direct schedule exploration,  providing a practical path from low-level execution evidence to reproducible, invariant-violating behaviors. 

\bibliographystyle{plain} 
\bibliography{sample-base}      
\end{document}